\documentstyle[prc,preprint,aps,epsf]{revtex}
\begin{document}
\draft
\title {Reliable calculations of level densities using statistical 
spectroscopy.}
\author{Jameel-Un Nabi$^1$, Calvin W.~Johnson$^1$ and W.~Erich Ormand$^2$ }
\address{
$^1$Department of Physics and Astronomy\\
Louisiana State University,
Baton Rouge, LA 70803-4001\\
$^2$Lawrence Livermore National Laboratory\\
P.O. Box 808, Mail Stop L-414 \\
Livermore, CA 94551
}

\maketitle
\begin{abstract}
Nucleosynthesis calculations require nuclear level densities for hundreds or 
even thousands
 of nuclides. Ideally one would like to
constrain these level densities by microscopically motivated yet
computationally cheap models. A statistical approach suggests that low moments
of the Hamiltonian might be sufficient. Recently Zuker proposed a simple
combinatoric formula for level densities based upon the binomial distribution.
We rigorously test the binomial formula against full scale shell-model
diagonalization calculations for selected $sd$- and $pf$-shell nuclides and also 
against Monte Carlo path integration calculations. We
find that the fourth moment is as important as the third moment to a good
description of the level density, as well as partitioning of the model space 
into subspaces.
\end{abstract}
\clearpage
\section {Introduction}
Nuclear level densities are important for the theoretical estimates of nuclear
reaction rates in
nucleosynthesis. The neutron-capture cross sections are approximately
proportional to the corresponding level densities around the nuclear resonance
region. The competition between neutron-capture and $\beta$ decay determines
the fate of
the \textit{s}- and \textit{r}-processes. The abundance of \textit{s}-process 
nuclei with nonmagic neutron number is inversely proportional to the 
neutron-capture cross section \cite{See65}, i.e., proportional to the level 
density. The waiting points of the \textit{r}-process are determined by the 
balance between the rates of neutron-capture and of photoejection of a neutron.

Reliable estimates of nuclear abundances require accurate level
densities. Authors like Gilbert and Cameron \cite{Gil65} gave
uncertainties beyond a factor of 10 in their early calculations and
uncertainties of about a factor of 8 and 5 are present in calculations of
\cite{Woo78} and \cite{Thi87}, respectively. What is desired is a reliability
within a factor of two.

At low excitation energies (up to 5 MeV), level densities are usually extracted 
experimentally via direct counting from neutron/proton resonance data 
\cite{Dil73}. For intermediate energies (between 5 and 15 MeV) they are 
extracted from charged particles spectra \cite{Hui69,Lu72} while for still 
higher energies the level densities are determined using the Ericson fluctuation 
analysis \cite{Hui69}. Recently, the Oslo group has reported on a new method to 
extract level density from primary $\gamma$-ray spectra without the assumption 
of any model \cite{Sch00}.

Such experimental measurements are very labor intensive, and reaction network 
calculations require hundreds or thousands of cross sections,
sometimes for unstable nuclides.  Therefore one turns to theoretical models
for help.

Most conventional calculations of the nuclear level density are
based on the Fermi gas model. The most widely used description of
the nuclear level density is the Bethe formula, based on a gas of
free nucleon  \cite{Bet36}. This parameterization works well for
the level densities of many nuclides at low energies and in the
neutron energy region  \cite{Boh69} especially when one uses the
modified ``backshifted Bethe formula'' \cite{Hol76,Cow91}.
Unfortunately the fitted single-particle level density parameter
and the backshift parameter are nucleus-dependent and are not
derived theoretically. Rauscher and collaborators presented a
global parametrization of nuclear level densities within the
back-shifted Fermi-gas formalism \cite{Rau97}. They employed an
energy-dependent level density parameter and a backshift parameter
to attain an improved fit of level densities at neutron-separation
energies. Despite all this, the fact remains that the input
parameters are just that: empirically fitted parameters.

Can one look to theory for help?  The interacting shell model and
other microscopic models accurately describe spectra and
transition for a broad range of nuclides.  On the other hand,
``traditional'' shell-model codes diagonalize the Hamiltonian in a
large-dimensioned basis of occupation-state wavefunctions but use
the Lanczos algorithm to  extract only a handful low-lying states.
The level density requires  all eigenstates and thus complete
diagonalization, a computationally forbidding requirement.

An alternative to diagonalization is the Monte Carlo path-integral
technique \cite{Joh92},
which is well suited to thermal observables \cite{Dea95,Nak98,Whi00}.
 Nakada
and Alhassid \cite{Nak98} extracted both the single-particle level
density parameter and the backshift parameter from the microscopic
Monte Carlo densities by fitting them to the backshifted Bethe
formula and also found some shell effects in their systematics.
Recent developments include  the particle number reprojection
method \cite{Has99}, which can  calculate level densities of
odd-even and odd-odd nuclei  despite a sign problem introduced by
the projection of an odd number of particles.  Path-integral
methods are limited, however, to interactions that are free of the
`sign problem.' Therefore we consider  alternatives.

One possibility is nuclear statistical spectroscopy
\cite{Fre71,Won86}, which argues that many nuclear properties are
controlled by low-lowing moments of the Hamiltonian.  Given a set
of one- and two-body interaction matrix elements (the same input
for a shell-model diagonalization or path-integral calculation)
one can readily find in the literature straightforward formulae
for first through fourth moments  \cite{Fre71,Won86,Ayi74}. One
early result of statistical spectroscopy was that for a finite
model space, the total level density tends to be Gaussian
\cite{Mon75}. Of course this means the model ignores intruders at
high excitation energy, but it also suggests that the level
density even at low energy is governed by the width, or second
moment, of the Hamiltonian. Grimes and collaborators included 4th
moment \cite{Gri79,Gri83} in order to improve the description of
the level densities. Recently Zuker  suggested a modified approach
\cite{Zuk99}. He argued that a binomial distribution rather than
Gaussian arises most naturally from combinatorial arguments.   In
the limit of an infinite number of states the binomial becomes a
Gaussian; but for a finite model space the differences are
nontrivial as we shall discuss.

This paper discusses in detail the implementation of a novel
treatment of level density descriptions using statistical models
introduced recently by Johnson, Nabi and Ormand \cite{Joh01}. Here
we compare these models against a number of nuclei whose level
densities can be fully computed by diagonalization of a two-body
Hamiltonian in the $sd$- or $pf$-shells. We also compare
against Monte Carlo path
integration calculations for nuclides where direct diagonalization
is not possible. We find that in most cases it is the {\it fourth}
moment which distinguishes among the exact, Gaussian, and binomial
level densities, although we find cases in which the 3rd moment is
important.

Section 2 presents the formalism of our calculations. Section~3
shows some comparison of the Gaussian and binomial level densities
with the microscopic level densities. In section 4 we present a
curious finding, that there is a relation between the fourth
moment of individual particle configurations and collectivity,
which suggests a possible way to use statistical spectroscopy to
model the low-lying, non-statistical collective structure of the
nuclides. Here we partition the model space into subspaces, use
the binomial distribution to construct partial level densities and
finally sum them  to get the total level density. These improved
models depict a more realistic picture of the microscopic level
densities. We conclude and summarize in Section~5.

\section{Formalism}

Nuclear statistical spectroscopy \cite{Fre71,Won86}, championed by
French and  collaborators, is built upon  moments of the nuclear
Hamiltonian $\hat{H}$ in a finite many-body space.  The first
moment or centroid is
\begin{equation}
\mu_{1} = \bar{H} \equiv \left \langle \hat{H} \right \rangle;
\end{equation}
all other moments are central moments:
\begin{equation}
\mu_n \equiv \left \langle \left (\hat{H}-\bar{H} \right )^n \right \rangle.
\end{equation}
As one computes higher and higher moments, one naturally regains
more and more information about $\hat{H}$.  But higher moments are
difficult to calculate.   An alternative is to partition the
model space
into subspaces ${\cal S}_\alpha$, using projection operators
$P_\alpha = \sum_{a \in \alpha} \left | a \right \rangle
\left \langle a \right | $.  
We call
\begin{equation}
\mu_n(\alpha) \equiv \left \langle
P_\alpha \left (\hat{H}-\bar{H} \right )^n \right \rangle
\end{equation}
a {\it configuration} moment, and general formulae can be found in
the literature for configuration moments up to $n=4$
\cite{Fre71,Ayi74}.  The total
central moments can be easily found by summing over the
configuration moments.

The density of states can be formally written as
\begin{equation}
\rho(E) = {\rm tr \,} \delta \left ( E - \hat{H} \right).
\end{equation}
One can also introduce {\it configuration} densities (sometimes
called {\it partial} densities)
\begin{equation}
\rho_\alpha(E) = {\rm tr \,} P_\alpha \delta \left ( E - \hat{H} \right),
\end{equation}
to which we will return  later.
If the {\it many-body} matrix elements of $\hat{H}$ are randomly
distributed (specifically, if they belong to a Gaussian Orthogonal Ensemble)
then the density of states of $\hat{H}$ follow a semi-circle
distribution.  If $\hat{H}$ however is a two-body operator then
the many-body matrix elements are correlated and the density of
states is Gaussian, or at least nearly so \cite{Mon75}.

The Gaussian and semi-circle distributions differ in their higher moments.
Let the {\it scaled} moments be defined by
\begin{equation}
m_n \equiv { \mu_n / (\mu_2)^{n/2} }, \hspace{0.8in} (n > 2),
\end{equation}
as the width $\mu_2$ provides a natural energy scale.  The scaled
fourth moment $m_4$ is 3 for a Gaussian but is 2 for the semi-circle
distribution. A Gaussian is reasonable good starting point as
it is already close to the actual level density.
A common generalization is to expand in a Gram-Charlier series
using Hermite polynomials \cite{Won86}; this unfortunately
can lead to negative level densities.  Another generalization is to
extend the Gaussian to a function of the form
$\exp(- \alpha E^2 - \beta E^3- \gamma E^4 \ldots ) $
\cite{Cha72,Gri83} but the relation
between  parameters
$\alpha , \beta , \gamma $ and the moments is not amenable to
a simple analytic formula.

Recently Zuker gave a combinatorial argument that
one should use a binomial distribution rather than  Gaussian
to approximate level densities \cite{Zuk99}.   We only summarize
his approach here.
Consider the binomial expansion
\begin{equation}
(1+\lambda)^N = \sum_{k = 0}^N \lambda^k.
{N \choose k}
\end{equation}
Now interpret this binomial expansion as a distribution of levels,
specifically, that
$\lambda^k {N \choose k}$ is the number of states at
excitation energy $E_x = \epsilon k$, $\epsilon$ being an
overall energy scale.
Because we can write ${N \choose k}$ with gamma functions, one
can easily generalize to the continuous case,
\begin{equation}
\rho(E_x) = \lambda^{E_x / \epsilon}
{ \Gamma(E_{max}/\epsilon+1) \over \Gamma(E_x / \epsilon +1)
\Gamma( (E_{max} - E_x) /\epsilon +1 )}
\end{equation}
where $E_{max} = \epsilon N$. Although we began with $N$ as an integer,
it no longer has to be.

The binomial distribution has several advantages.  In the limit
$\lambda =1, N\rightarrow \infty$ one regains the Gaussian. For
$\lambda \neq 1$ the distribution is asymmetric, because of a
non-zero third moment; Zuker makes a combinatorial argument that
the third moment can be as significant as the second moment.
Finally, unlike most generalizations to the Gaussian, such as
adding a $\gamma E^4$ term to the argument of the exponential, one
can easily compute the moments of the binomial. The total number
of levels, which is in effect the `zeroth' moment, is
\begin{equation}
d= (1+\lambda)^N,
\label{moment0}
\end{equation}
while the centroid and
width are given by
\begin{eqnarray}
\mu_1 = {N \epsilon \lambda \over 1+\lambda} \\
\mu_2 = {N \epsilon^{2} \lambda \over (1+ \lambda)^{2}}
\end{eqnarray}
and the {\it scaled} third and fourth moments are
\begin{eqnarray}
m_{3}={1- \lambda \over \sqrt{N \lambda}},
\label{moment3}\\
m_4 =
3- {4-\lambda \over N}+ {1 \over N \lambda}.
\label{moment4}
\end{eqnarray}
Ref \cite{Zuk99} does not give the fourth central moment.
For Gaussians ($N \rightarrow \infty $) $m_4=3$. For
most binomials and for shell model diagonalization, the scaled fourth moment
is less than 3, a typical value being around 2.8. (Note that the above moments 
are exact for
discrete distributions but are only approximate for the
continuous distributions, due to errors in integration.
For large $N$, however, they are very good approximations.)

Using Stirling's approximation, and a few others,
Zuker arrives at
\begin{equation}
\rho(E_{x}) \approx \sqrt{\frac{8}{N \pi}}
\exp \left (
-(N-1)\left (\frac{E_{x}}{E_{max}}\ln{\frac{E_{x}}{E_{max}}}+
\frac{E_{max}-E_{x}}{E_{max}}\ln{\frac{E_{max}-E_{x}}{E_{max}}}\right
)+N\frac{E_{x}}{
E_{max}}\ln{\lambda} \right ).
\end{equation}

The key parameters of the binomial are the order $N$ and the
asymmetry parameter $\lambda$.  If $\lambda=1$ then the binomial
is {\it symmetric} and has $m_3=0$; if $\lambda \neq 1$ then the
binomial is {\it asymmetric} or skewed.  The skewness can be
positive ($m_3 > 0, 0 < \lambda < 1$), or negative  ($m_3 < 0,
\lambda > 1 $). Zuker suggested that the order of the binomial,
$N$, be fixed by the dimension of the model space. In that case
$N$ and $\lambda$ are fixed by solving eqns. (\ref{moment0}) and
(\ref{moment3}) simultaneously.  This we consider to be the
`standard' binomial.  We note, however, that one could instead fix
the order $N$ by the fourth moment, and solve (\ref{moment3}) and
(\ref{moment4}) simultaneously instead, afterwards multiplying the
entire binomial distribution by a constant so as to get the
correct total number of levels. This we refer to as the {\it
fourth-moment scaled} (FMS) binomial. After $N$ and $\lambda$ are
determined, the centroid and width simply fix the absolute scale
of the distribution.

The central question of this paper is which distribution--Gaussian, binomial,
FMS binomial, or some (other) improved model--best describes microscopic level 
densities?  We use the low
moments, in particular the third and fourth moment, to characterize
our results.

\section{Comparison of level densities without partitions}

To test binomials as candidates for modeling level densities, 
we compare against exact  shell model calculations. 
(All the densities shown in our figures are in fact {\it state} density, which 
includes  $2J+1$ degeneracies.  Strictly speaking, the {\it level
density} ignores $2J+1$ degeneracies, but the literature is often cavalier
with this distinction).
In this section we do not partition the model space into configuration 
subspaces, which will be considered in the next section. 

We considered full $0\hbar\omega$ $sd$- and $pf$-shell calculations,
and focused on  several nuclides  that could
be completely diagonalized using the OXBASH shell-model code \cite{oxbash}.
In the $sd$-shell we used the Wildenthal USD
interaction \cite{Wil84} and performed exact shell model
calculations for many nuclides covering the entire range. In the
$pf$-shell we used the Brown-Richter interaction \cite{richter}
and computed $^{44,45}$Ti. 
For still larger dimensions, we turned to Monte
Carlo path integration calculations
\cite{Joh92,Dea95,Nak98,Whi00}. To avoid the well-known sign
problem \cite{Joh92} we fitted a schematic multipole-multipole
interaction  to the $T=1$ matrix elements of the Brown-Richter 
interaction \cite{richter}. For all of these calculations we
computed the exact spectroscopic moments, using the same
single-particle energies and two-body interaction matrix elements
were used by OXBASH.  As far as we know this is the first time the
exact formulas of \cite{Ayi74} for third and fourth moments have
been computed for a general case \cite{code}.

We ask: Is there a significant difference between
exact, Gaussian, and binomial level densities, and is the difference
driven by third or fourth moments? First the moments.
Table~1 shows the scaled third and fourth moments for the exact
distributions (the moments from exact shell model
eigenvalues and those calculated using \cite{Ayi74,code} should and
do agree) as well as fourth moment for
`standard' binomials for some selected $sd$- and $pf$-shell nuclides.  
(Remember that the third and fourth moments
for a Gaussian are 0 and 3, respectively.
)
 The third moment is generally small, except for
the case of $^{20}$F, $^{24}$Ne, $^{28}$Na and $^{36}$Ar in the $sd$-shell and 
$^{54}$Mn and $^{54}$Fe in the $pf$-shell; we found a strong correlation between
nontrivial $m_3$ and $T_z \neq 0$.
Because the standard binomial parameters $N$ and $\lambda$
are found by solving
Eqn. (\ref{moment0}) and (\ref{moment3}) (that is, constrained by
the total number of levels and the third moment ), the
fourth moment, $m_4$,
is {\it not} constrained.
We see from Table~1 that on the average, the (standard) binomials tend to 
overestimate the scaled fourth moment, $m_4$. We further note that whenever the 
magnitude of the scaled third moment is $\geq 0.05$ (which we regard as a 
significant value), the difference in the exact and binomial scaled fourth 
moment is considerable, albeit less so for the $pf$-shell nuclides.

Fig.~1 shows two sample $sd$- and two $pf$-shell nuclides. The histograms are 
the result of direct diagonalization. The figure shows the exact spectrum, the 
Gaussian and the two binomials (symmetric and asymmetric). 
For the symmetric case we forced the asymmetry parameter $\lambda$ to
be 1 (the fourth moment did not change significantly).
Because the symmetric and asymmetric binomial distributions are very
similar for all these cases, we conclude that it is the fourth moment rather 
than the third
moment that drives the difference.  For $^{45}$Ti, in fact
the symmetric and asymmetric binomials are indistinguishable. We remind the 
reader that our goal is a factor of 2 accuracy in modeling the level density.

But what if the scaled third moment is significantly different from zero (i.e. 
if the absolute value is $\geq 0.05$)? Intuition dictates that the third moment 
should play a role for such cases.
Figure~2 shows the case for nuclides with large asymmetry. Again we chose two  
$sd$- and two  $pf$-shell nuclides.
 The open circles in Fig.~2 show the Monte Carlo shell model calculations along 
with the uncertainties. Here there is a clear difference between the symmetric 
and asymmetric binomials as expected.
For these cases there also exist a large discrepancy between the binomial and 
exact $m_4$.  We therefore also plot the FMS (fourth-moment-scaled) binomial,
which has the correct $m_4$. The FMS binomial has clearly a better
behavior especially for the $pf$-shell nuclides. Unfortunately the scaling also 
leads to a cut-off (unlike
a Gaussian, the binomial has sharp cutoffs and  infinite slope at
$E_x=0,E_{max}$) which is too high for the case of  $^{24}$Ne.

The preliminary lessons that we learn:  binomial distribution {\it do}
model shell-model level densities better than Gaussians, but in many
cases it is the fourth moment rather than the third that leads to a
discrepancy.  Nuclides with $T_3 \neq 0$ have larger asymmetries,
however, and must be accounted for.
A scaled third moment of about $\pm 0.05$ is non-negligible.
While an obvious solution is to adjust the binomial to reproduce the
exact fourth moment, the binomial's abrupt cut-off in certain cases, 
leads us to generalize our approached to a partitioned model space, 
described in the next section. 

\section{Sum of Partitioned Binomial (SUPARB) level densities}

To improve our results we partition the model space into 
configuration subspaces.
One expects partitioned models to better approximate the higher moments 
\cite{Plu88,kota} (for example, Pluhar and Weidenm\"uller \cite{Plu88} had a sum 
of distorted semi-circles and got much better modeling than a single 
semi-circle). We also saw in the previous section that the FMS binomial gets the 
fouth moment correct but stops abruptly. Partitioning of the model space into 
subspaces solves this problem.

We partitioned the model space by single-particle configurations, 
such as $(1s_{1/2})^2 (0d_{3/2})^1 (0d_{5/2})^3$,
 for the very simple reason that
analytic formulas exist for the associated configuration moments.
Other possibilities would be to partition by $J, T$ or by SU(3);
unfortunately no analytic formulas exist for these partitions.
We did consider $J, T$ partitions, computing the moments by hand
(the OXBASH shell model code projects the Hamiltonian onto
$J, T$ partitions) but we did not find significant advantage.  SU(3)
would be an attractive partitioning , as it would naturally build in quadrupole 
collectivity, but even by hand this is
difficult for a general two-body interaction.
Experts should note that protons and neutrons 
orbits were considered distinct, which would allow 
us to easily project exact isospin. 

To compute the level densities, we take the following steps: (1) We compute the 
configuration moments up to 4th order. (2) We model the partial density for each 
configuration as a binomial (either standard or FMS). The binomial parameters 
$N_\alpha$ and $\lambda_\alpha$, as well as the overall energy scale and 
centroid, are fitted to the configuration moments for the $\alpha^{th}$ 
subspace. The binomial then gives a partial density. (3) The partial densities 
are summed to yield the total level densities. Because of these ingredients, we 
will refer to our approach as SUPARB (SUm of PARtitioned Binomials) level 
densities. If we model the partial densities for the configurations as a FMS 
binomial, i.e., if we fix the order $N_\alpha$ by the fourth moment,
and solve (\ref{moment3}) and (\ref{moment4}) simultaneously for each 
configuration, we call the resulting sum as SUPARB-FMS.

Fig.~3 models the state densities for four $sd$-shell nuclides,  $^{23}$Ne,  
$^{24}$Mg, $^{32}$P, and $^{32}$S. Here we plot the SUPARB and SUPARB-FMS state 
densities against the exact ones. We see that not only the SUPARB-FMS densities 
model the exact state densities some what better, some structure in the 
low-lying states is also revealed. Such structure cannot be described
by simple models such as either a Gaussian or a binomial. The structure is 
particularly very profound for the low-lying states of $^{24}$Mg and $^{32}$S. 
There were however cases where no appreciable difference between the SUPARB and 
SUPARB-FMS state densities was found. Fig.~4 shows four such cases: 
$^{22}$Ne,  $^{22,24}$Na, and  $^{23}$Mg.
 We see that the comparison is again very good (within a factor of 2) and that 
there is no appreciable difference between the SUPARB and SUPARB-FMS state 
densities.

The computation time of moments, $\mu_n$, scales as (number of 
partitions)(number of j-orbits)$^{2n}$. That is, calculation of fourth moments 
requires much more CPU time as compared to the third one. SUPARB-FMS is sometime 
a clearly superior model but there are also cases where SUPARB alone gives 
equally good results. Two important questions then arise:
(1) Is it possible to determine the cases where SUPARB state densities are 
sufficient, merely by looking at the configuration third moments in order to 
save time?
 and
(2) What are the reasons behind when SUPARB-FMS gives better results? When do we 
see the low-lying structure in SUPARB-FMS?

The answers to these questions do not seem to be simple.  There are clues,
however, to the choice between the two SUPARB densities. One clues lies in the 
spread of the scaled third configuration moments. If there is a spread of these 
moments both above 0.05 and below -0.05, SUPARB-FMS is likely to produce better 
results. Fig.~5 shows the spreads in the scaled third configuration moments 
$m_3(\alpha)$, where $\alpha$ labels the configuration, for some selected 
nuclides from Figs.~3 and 4.  The ordinate is the associated configuration 
centroid $h_\alpha$ which indicates the average excitation energy 
of states in the subspace $\alpha$. (We have neglected a few of the 
configuration third moments in all the four cases with values much less than 
-0.05.) The  bulk of the $m_{3}(\alpha)$ lie in the range $-0.2 \leq m_3 \leq 
0.2$. Any  $m_{3}(\alpha)$ which falls outside the rectangle is to be taken as 
significant. What we see is that for the graphs in the left column there is a 
spread of $m_3$ around the rectangle. For the right column we do not see any 
$m_3 >$ 0.05. And we see that for these two cases ($^{22}$Ne and  $^{24}$Na), 
standard SUPARB densities  do equally well (see Fig.~3). So for these cases 
there is no special need for computation of the scaled fourth moments (similar 
is the case for  $^{23}$Mg and  $^{22}$Na).

 There can be a problem for  high positive values of $m_3$,  
$\lambda \rightarrow 0$, and the binomial 
$m_3 \rightarrow \frac{1}{\sqrt{\ln d}}$. So, for example, 
if the dimension $d= 500$, by Equation (\ref{moment3})
the maximum $m_3$ allowed by the standard binomial is 0.16. For any value of 
exact $m_3> 0.16$ the standard binomial fails. Frequently we found 
large values of $m_3$ for configurations even of small dimension. 
Fortunately, for fourth-moment-scaled (FMS) binomials, this is not 
as limiting  (we fix the order $N_\alpha$ by the fourth moment,
solve (\ref{moment3}) and (\ref{moment4}) simultaneously, and
afterwards multiply the entire binomial distribution by a constant
so as to get the correct total number of levels) and the technique can be 
applied to all cases.

Regarding the answer to the second question, the reply lies partly in what we 
already described previously. Empirically we find that a spread of exact 
$m_3(\alpha)$ around $\pm 0.05$, is likely to reveal some structure. Again if we 
look at Fig.~5 we do see that for $^{23}$Ne and $^{32}$P there is a spread of 
$m_3$ above and below the rectangle; simultaneously, Fig.~4 shows 
nontrivial structure at very low excitation energy for these nuclides. 
The same pattern occurs for  $^{24}$Mg and $^{32}$S, suggesting that the two 
events are correlated.

Yet another clue lies in the  $m_{4}(\alpha)$ shown in Fig.~6. We note that for 
cases where SUPARB-FMS densities reveal some structure, the standard binomials 
systematically overestimate the configuration scaled fourth moments. One clearly 
sees that for  $^{24}$Mg and $^{32}$S, the (standard) binomial  $m_{4}(\alpha)$ 
are biased high compared to their exact values. On the other hand, for  
$^{22}$Na and $^{23}$Mg, relatively structureless at low excitation energy, 
the binomial  $m_{4}(\alpha)$'s are not biased relative to their  exact values
but fall both high and low symmetrically.

Let us summarize what we have said so far for these improved statistical models 
of level densities. SUPARB-FMS density can be a better choice, not only in 
modeling the secular behavior (within a factor of two) but also reveals some 
structure in certain cases, specially for the interesting low-lying region. It 
however, requires a larger CPU time which can be a big problem when modeling 
thousands of cases. SUPARB density is still better compared to a single binomial 
density but for large, positive $m_3$ (which is very common for partitioned 
subspaces) they are constrained by the dimension of the configurations.

\section{Conclusions}
We have outlined a theoretical approach to level densities
that is both microscopic in origin and also computationally tractable. 
 We analyzed our results in terms of the third and fourth moments. For small 
asymmetries, the fourth moment dominates. For cases where the scaled
third moment is significantly different from zero (with magnitude greater than
0.05) the binomials are characterized by both the third and fourth central
moments. The FMS binomial gives the best description of the exact level
densities, specially when the nuclides have a large scaled third moment. There
is an improvement over the asymmetric standard binomial by as much as a factor 
of 2 in
the modeling of the low-lying exact level densities. Our study also shows
that the Gaussians continue to give a good estimate of the exact spectrum for
the level densities (within a factor of 5) and this comes quite handy since
they just require the centroid and width of the Hamiltonian as input
parameters. They might be used for an initial estimate of level density and then 
the interesting cases can be followed by still refined models presented here.

Partitioning the space considerably improves the binomial approximation to the
level densities. The SUPARB-FMS has some in-built structure
and follows the trend of the exact level density specially in the collective
regime.  The degree of precision of the present approach will give astrophysical 
nucleosynthesis calculations a much better predictive power.
Future
application to higher shells will be  hampered as much by our ignorance of the
effective two-body interaction as anything else.
SUPARB can also be easily generalized to the calculation of spin-cutoff factors, 
and we are planning to make application to  
nuclear heat capacities
and estimate of contamination by spurious center-of-mass motion.

This work was performed under the auspices of the
Louisiana Board of Regents,
contract number LEQSF(1999-02)-RD-A-06; and under the auspices of the U.S. 
Department of Energy
through the University of California, Lawrence Livermore National Laboratory,
under contract No. W-7405-Eng-48.

\onecolumn
\textbf{Table (1):} Scaled third and fourth moments. The exact fourth moments 
are compared with those of standard binomials. \\ \\
\begin{tabular}{|c|c|c|c||c|c|c|c|}
\hline
Nuclides &  $m_3$ &  $ {m}_{4}$ (exact) & $m_4$ (standard & Nuclides &  $m_3$ &  
$ {m}_{4}$ (exact) & $m_4$ (standard \\
& & & binomial)& & & & binomial) \\ \hline
$^{20}$F &  0.12 & 2.60 &  2.86 & $^{32}$Si &  0.024 & 2.90  & 2.87 \\
$^{20}$Ne &  0.092 & 2.71 &  2.83 & $^{30}$P &  0.0053 & 2.79  & 2.89  \\
$^{24}$Ne &  0.10 & 2.80 &  2.92  & $^{32}$P &  -0.023 & 2.78  & 2.86 \\
$^{22}$Na & 0.036 & 2.81 &   2.87  & $^{32}$S &  -0.037 & 2.77 & 2.86  \\
$^{24}$Na & 0.057 & 2.78 &   2.90   & $^{30}$Cl &  0.054 & 2.76 & 2.90 \\
$^{26}$Na & 0.088 & 2.76 &   2.92   & $^{32}$Cl &  -0.021 & 2.84 & 2.86 \\
$^{28}$Na &  0.14 & 2.73 & 2.94  & $^{34}$Cl &  -0.073 & 2.88 & 2.82 \\
$^{23}$Mg &  0.035 & 2.80 &  2.88 & $^{36}$Ar &  -0.12 & 2.97 &  2.74 \\
$^{24}$Mg &  0.036 & 2.81  & 2.89 & $^{44}$Ti &  0.036 & 2.82 & 2.87 \\
$^{24}$Al &  0.058 & 2.83  & 2.90 & $^{45}$Ti &  -0.0026 & 2.89  & 2.87 \\
$^{26}$Al &  0.039 & 2.81  & 2.90 & $^{48}$Cr &  -0.071 & 2.96  & 2.89 \\
$^{28}$Al &  0.038 & 2.79  & 2.90 &  $^{54}$Mn &  -0.11 & 2.96 & 2.91\\
$^{30}$Al &  0.054 & 2.77  & 2.90 & $^{54}$Fe &  -0.11 & 2.93  & 2.91  \\
\hline
\end{tabular}

\begin{figure}
\epsfxsize=11.0cm
\epsffile{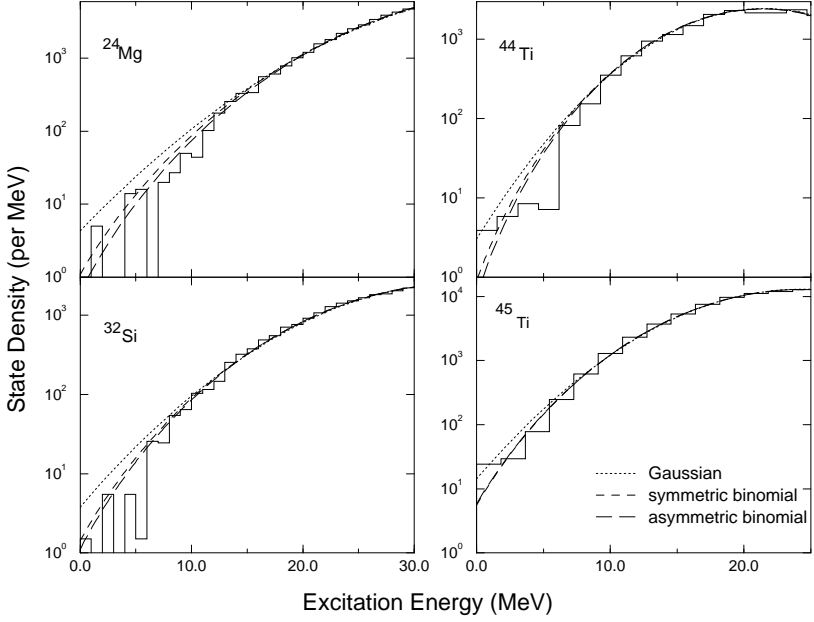}
\caption{ \footnotesize Comparison of exact shell model state densities 
(histograms) against Gaussians and binomials for $sd$- and $pf$-shell nuclides.}
\end{figure}

\begin{figure}
\epsfxsize=11.0cm
\epsffile{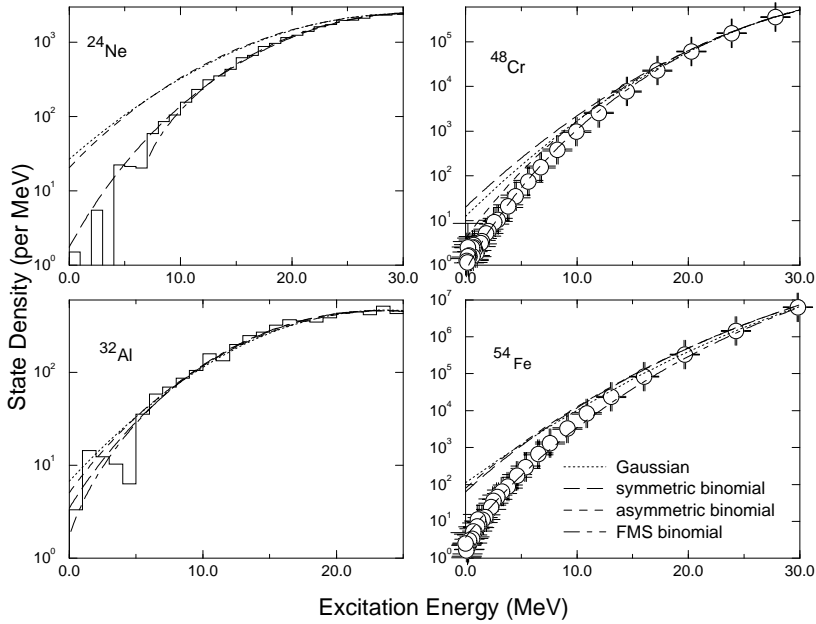}
\caption{ \footnotesize Comparison of exact shell model state densities 
(histograms) against Gaussians and binomials for $sd$- and $pf$-shell nuclides. 
The circles represent the Monte Carlo path integration calculation in a full 
$0\hbar\omega$ basis along with the uncertainties. FMS = ``fourth moment 
scaled'' (see
text).}
\end{figure}

\begin{figure}
\epsfxsize=12.0cm
\epsffile{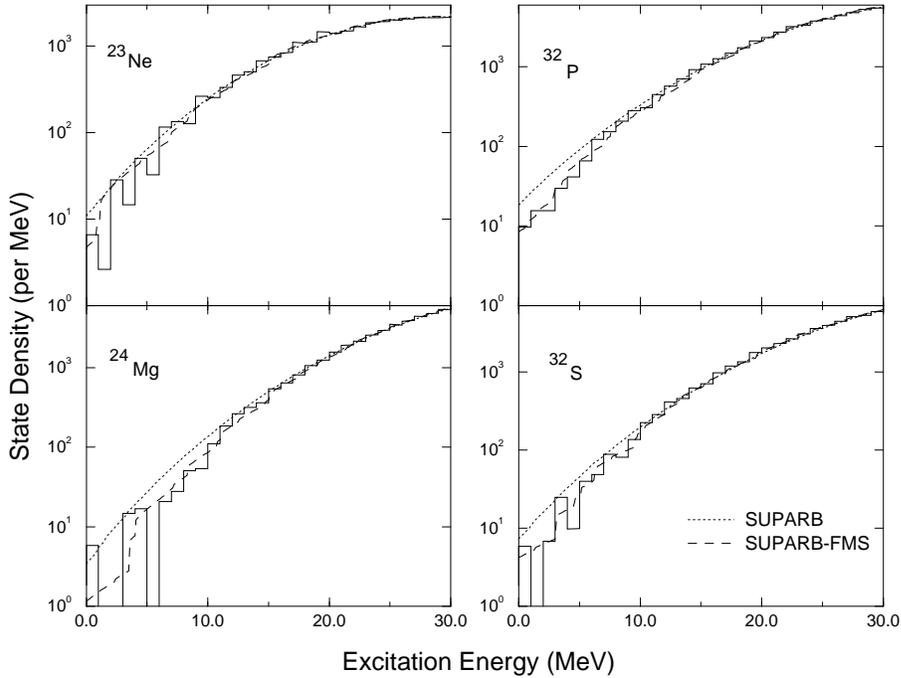}
\caption{ \footnotesize Comparison of exact shell model state densities 
(histograms) against sum of partitioned binomial densities (SUPARB) and sum of 
partitioned binomial densities which are fourth moment scaled (SUPARB-FMS). Note 
the structure in the SUPARB-FMS densities.}
\end{figure}

\begin{figure}
\epsfxsize=12.0cm
\epsffile{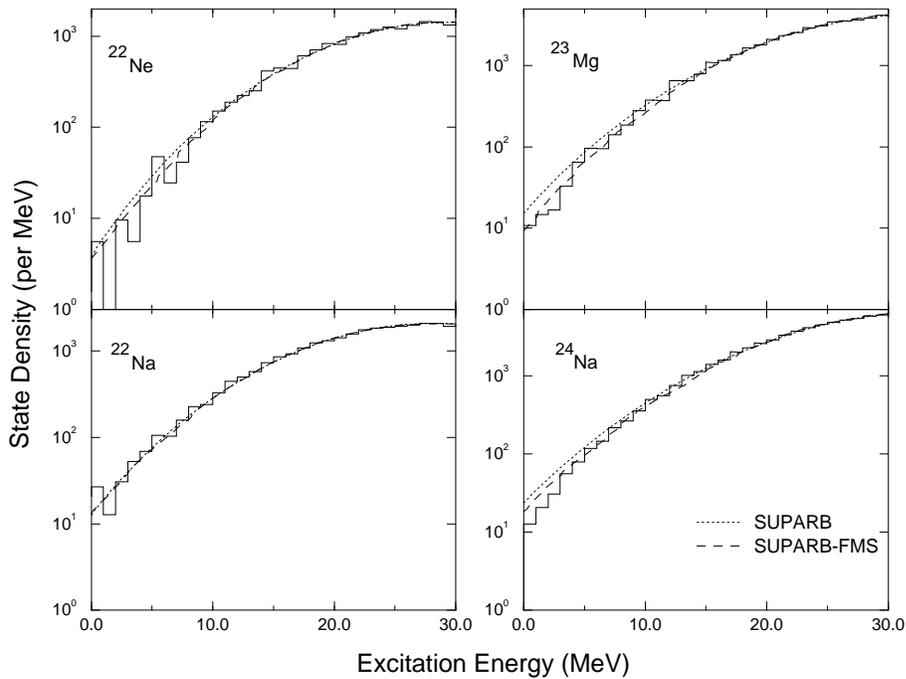}
\caption{ \footnotesize Same as Fig.~3 but for four other $sd$-shell nuclides. 
Here there is little distinction between standard and fourth-moment-scaled 
binomials.}
\end{figure}

\begin{figure}
\epsfxsize=11.0cm
\epsffile{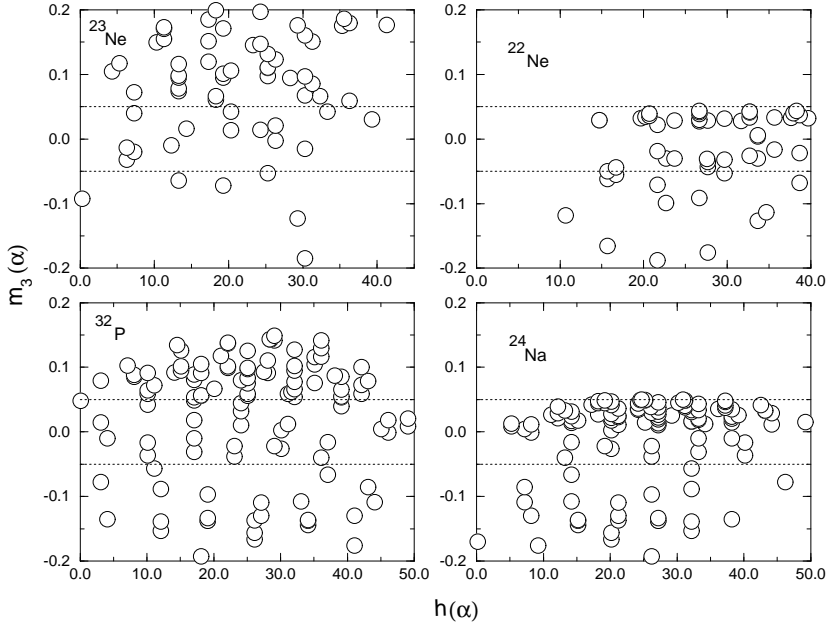}
\caption{ \footnotesize Spread of configuration scaled third moments 
$m_{3}(\alpha)$ as a function of configuration centroids $h(\alpha)$ 
(relative to the ground state energy). Any value of $m_{3}(\alpha)$ outside the 
dotted rectangle represents a significantly high value.}
\end{figure}

\begin{figure}
\epsfxsize=11.0cm
\epsffile{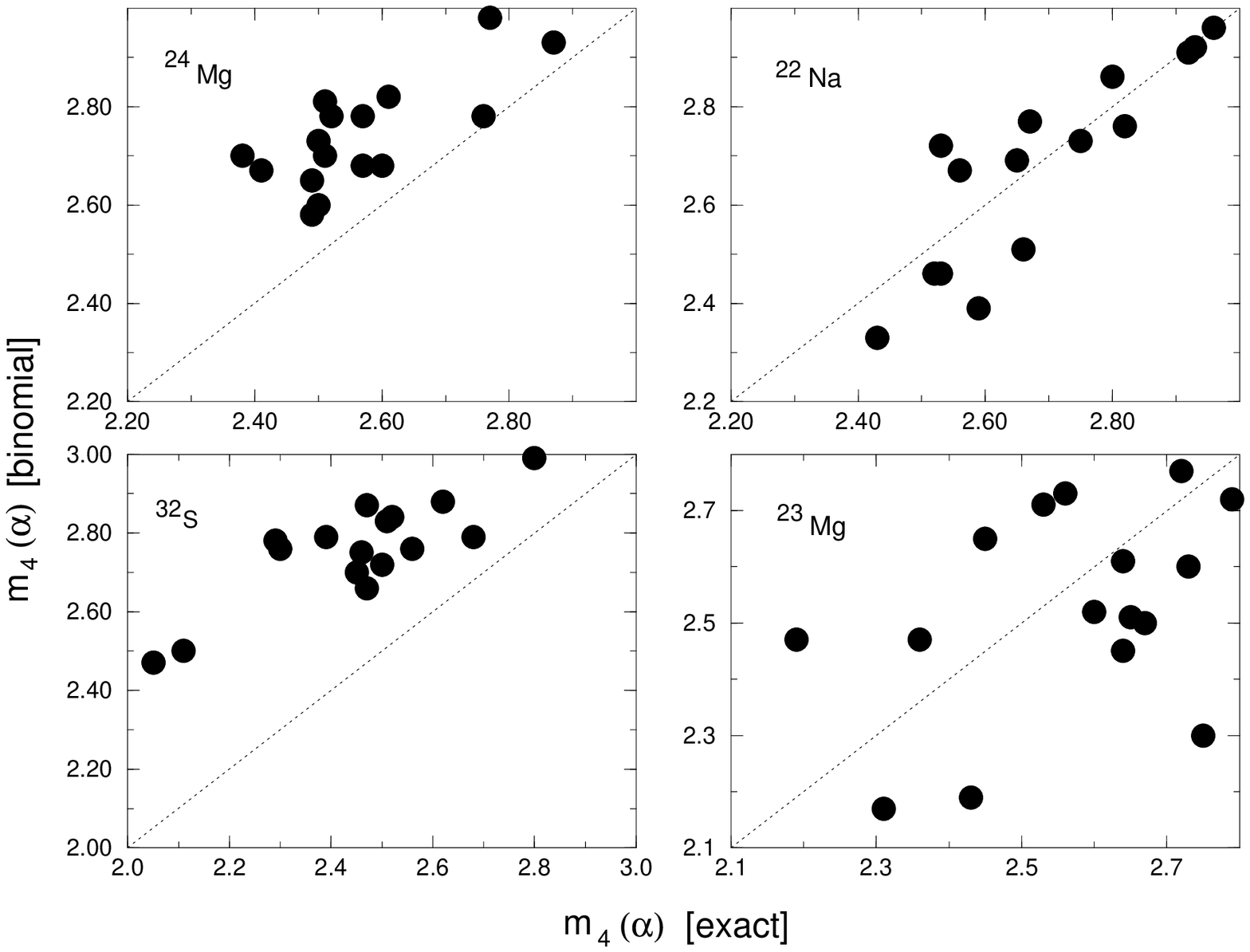}
\caption{ \footnotesize Comparison of the low-lying exact (configuration) scaled 
fourth moments $m_{4}(\alpha)$ with those calculated from the standard binomial. 
The dotted line is to guide the eye.}
\end{figure}


\begin{thebibliography}{99}
\bibitem{See65}P.~A.~Seeger, W.~A.~Fowler, and D.~D.~Clayton, Astrophys. J. 
Suppl. {\bf 11}, 121 (1965).
\bibitem{Gil65}A. Gilbert and A.G.W. Cameron, Can. J. Phys. {\bf 43}, 1446
(1965).
\bibitem{Woo78}S.E. Woosley, W.A. Fowler, J.A. Holmes and B.A. Zimmerman, At.
Nucl. Data Tables {\bf 22}, 371 (1978).
\bibitem{Thi87}F.-K. Thielemann, M. Arnould and J.W. Truran, Advances in
Nuclear Astrophysics, eds. E. Vangioni-Flam et. al. p.525.
\bibitem{Dil73}W. Dil, W. Schantl, H. Vonach and M. Uhl, Nucl. Phys. A {\bf
217}, 269 (1973).
\bibitem{Hui69}J.R. Huizenga, H.K. Vonach, A.A. Katsanos, A.J. Gorski and C.J.
Stephan, Phys. Rev. {\bf 182},1149 (1969).
\bibitem{Lu72}C.C. Lu, L.C. Vaz and J.R. Huizenga, Nucl. Phys. A {\bf 190},
229 (1972).
\bibitem{Sch00}A. Schiller, L. Bergholt, M. Guttormsen, E. Melby, J. Reststad
and S. Seim, Nucl. Instum. Methods A {\bf 447}, 498 (2000).
\bibitem{Bet36}H.A. Bethe, Phys. Rev. {\bf 50}, 332 (1936).
\bibitem{Boh69}A. Bohr and B.R. Mottelson, Nuclear Structure vol. 1 (Benjamin,
New York, 1969).
\bibitem{Hol76}J.A. Holmes, S.E. Woosley, W.A. Fowler and B.A. Zimmerman,
Atom. Data, Nucl. Data Tables {\bf 18}, 305 (1976).
\bibitem{Cow91}J.J. Cowan, F.-K. Thielemann and J.W. Truran, Phys. Rep. {\bf
208}, 267 (1991).
\bibitem{Rau97}T. Rauscher, F.-K. Thielemann and K.-L. Kratz, Phys. Rev. C
{\bf 56}, 1613 (1997).
\bibitem{Joh92}C.W. Johnson, S.E. Koonin, G.H. Lang and W.E. Ormand, Phys.
Rev. Lett/ {\bf 69}, 3157 (1992).
\bibitem{Dea95}D.J. Dean, S.E. Koonin, K. Langanke, P.B. Radha and Y.
Alhassid, Phys. Rev. Lett. {\bf 74}, 2909 (1995).
\bibitem{Nak98}H. Nakada and Y. Alhassid, Phys. Lett. B {\bf 436}, 231 (1998).
\bibitem{Whi00}J.A. White, S.E. Koonin and D.J. Dean, Phys. Rev. C {\bf 61},
034303 (2000).
\bibitem{Has99}Y. Alhassid, S. Liu and H. Nakada, Phys. Rev. Lett. {\bf 83},
4265 (1999).
\bibitem{Fre71}J.B. French and K.F. Ratcliff, Phys. Rev. C {\bf 3}, 94 (1971).
\bibitem{Won86}  S.~S.~M.~Wong, {\it Nuclear Statistical Spectroscopy},
Oxford Press (New York, 1986).
\bibitem{Ayi74}S. Ayik and J.N. Ginocchio, Nucl. Phys. A {\bf 221}, 285 (1974).
\bibitem{Mon75} K. K. Mon and J. B. French, Ann. Phys. {\bf 95}, 90 (1975).
\bibitem{Gri79}S. M. Grimes, S. D. Bloom, R. F. Hausman, Jr. and B. J. Dalton, 
Phys. Rev. C {\bf 19}, 2378 (1979).
\bibitem{Gri83}S. M. Grimes, S. D. Bloom, H. K. Vonach and R. F. Hausman, Jr., 
Phys. Rev. C {\bf 27}, 2893 (1983).
\bibitem{Zuk99}A. P. Zuker,  Phys. Rev. C {\bf 64}, 021303 (2001).
\bibitem{Joh01}C. W. Johnson, J.-U. Nabi and W.~E.~Ormand, submitted to Phys. 
Rev. Lett. (2001), LANL archive nucl-th/0105041.

\bibitem{Cha72}F. S. Chang and A. Zuker, Nucl. Phys. A {\bf 198}, 417 (1972).
\bibitem{oxbash} B.~A.~Brown, A.~Etchegoyen, and W.~D.~M.~Rae, OXBASH,
the Oxford University-Buenos Aires-MSU shell model code, Michigan State
University Cyclotron Laboratory Report No. 524 (1985).
\bibitem{Wil84}
B.H. Wildenthal, Prog. Part. Nucl. Phys. {\bf 11}, 5 (1984).
\bibitem{richter} W.~A.~Richter, M.~G. van der Merwe, R.~E.~Julies, and
B.~A.~Brown, Nucl. Phys. {\bf A523}, 325 (1991),  Nucl. Phys. {\bf A557}, 585 
(1994).
\bibitem{code}J.-U. Nabi and C.~W. Johnson, program CONMOM4, to be published.
\bibitem{Plu88} Z.~Pluhar and H.~A.~Weidenm\"uller, {\it Phys.~Rev.~C }
{\bf 38} (1988)  1046.
\bibitem{kota}V.~K.~B.~Kota, D.~Majumdar, Nucl. Phys. {\bf A604}, 129 (1996).
\end{thebibliography}
\end{document}